\documentclass[%
reprint,
amsmath,amssymb,
aps,
prl,
]{revtex4-1}

\usepackage{graphicx}
\usepackage{amsmath}
\usepackage{amssymb}
\usepackage{slashed}
\usepackage{bbm,bm}
\usepackage{hyperref}
\usepackage{xcolor}
\usepackage{comment}
\usepackage{amsfonts}
\usepackage{dsfont}
\usepackage{soul}

\def\bea{\begin{eqnarray}} 
\def\eea{\end{eqnarray}}
\def\be{\begin{equation}} 
\def\ee{\end{equation}} 
\def\ba{\begin{array}}
\def\ea{\end{array}}

\def\be{\begin{equation}}
\def\ee{\end{equation}}
\def\bea{\begin{eqnarray}}
\def\eea{\end{eqnarray}}

\begin{document}

\title{Are there any problem for large values of the action like there were
for small ones?}

\author{Ennio Gozzi}
\email{ennio.gozzi@gmail.com}
\affiliation{INFN, Section of Trieste, Via Valerio 2, Trieste, 34100, Italy, and Phys. Dept. Theoretical Section, Univ. of Trieste,
Strada Costiera 11, Miramare, Grignano, Trieste, 34152, Italy}

\begin{abstract}
In this paper we show that a method of quantization proposed few years
ago (Ann.~of Physics (314) 2005, 24) is equivalent to studying the
system for values of the action close to zero. In this paper  we also
study the behaviour when the action gets very very large which could
be the regime where dark energy and dark matter are invoked. 
\end{abstract}

\maketitle

\section{Brief review}

In the 30's Koopman and von Neumann \cite{KvN} proposed an Hilbert space
and operatorial approach to classical mechanics (CM). In the 80's
and 90's this approach was turned into a path integral. All this work
is summerized in ref.~\cite{AGM} to which we will refer the reader for
more details. In the same paper we showed how quantization can be
achieved in this framework.

Skipping all the details, the generating functional for this path
integral is given by \cite{AGM}:
\begin{eqnarray}
Z & = & \int{\cal D}\Phi\,\exp\left[i\int dt\,d\theta\,d\bar{\theta}\,iL\left(\Phi\right)\right]\,,\label{eq:def-generating-functional}
\end{eqnarray}
where $L$ is the usual Lagrangian of the system, $\theta$ and $\bar{\theta}$
are two Grassmannian partners of time $t$, the $\Phi$ are superfield
extensions of the phase space $\left(q,p\right)$ of the system. These
extensions are defined by
\begin{eqnarray}
\Phi^{a}\left(t,\theta,\bar{\theta}\right) & = & \varphi^{a}\left(t\right)+\theta c^{a}\left(t\right)+\bar{\theta}\omega^{ab}\bar{c}_{b}\left(t\right)+i\bar{\theta}\theta\omega^{ab}\lambda_{b}\left(t\right)\,, \nonumber \\
&&\label{eq:def-superfield}
\end{eqnarray}
where $\varphi^{a}$ are the phase space variables of the system,
$\omega^{ab}$ the symplectic matrix of the Hamiltonian equations
of motion, and $c^{a}$, $\bar{c}_{a}$, $\lambda_{a}$ are auxiliary
variables whose geometrical nature has been clarified in several old
papers quoted in ref.~\cite{AGM}.

The quantization of the system is achieved by multiplying $L\left(\Phi^{a}\right)$
by $\frac{\theta\bar{\theta}}{\hbar}$. In fact, as proved in ref.~\cite{AGM},
this procedure brings the classical generating functional (\ref{eq:def-generating-functional})
to the following one:
\begin{eqnarray}
Z & = & {\cal N}\int{\cal D}\varphi\,\exp\left[\frac{i}{\hbar}\int dt\,L\left(\varphi\right)\right]\,,\label{eq:def-QM-generating-functional}
\end{eqnarray}
which is the generating functional of the associated quantum system
(QM) modulo a normalization constant ${\cal N}$. In ref.~\cite{AGM}
we studied the geometrical meaning of this procedure. We proved that
it is equivalent to ``geometric quantization'' \cite{woodhouse-book}.

In this paper we shall now study the physical meaning of the procedure above.

\section{Small values of the action \label{sec:Small-values-of-S}}

Let us first do a dimensional analysis. The argument of the exponential
in (\ref{eq:def-generating-functional}) is just a phase. So $d\theta\,d\bar{\theta}$
has the dimension of the inverse of the action $\int dt\,L\left(\Phi\right)$.
Next let us notice that the factor $\theta\bar{\theta}$ by which
we multiply the $L\left(\Phi\right)$, in order to get quantum mechanics
(modulo $\hbar$), is equivalent to $\delta\left(\bar{\theta}\right)\delta\left(\theta\right)$,
which has the dimension of the inverse of $d\theta\,d\bar{\theta}$
because $\int d\theta\,d\bar{\theta}\,\delta\left(\bar{\theta}\right)\delta\left(\theta\right)=1$.
As a consequence   $\bar{\theta}\theta$ has the dimension of an action. Multiplying
by this factor, as it is equivalent to $\delta\left(\bar{\theta}\right)\delta\left(\theta\right)$,
means choosing values of the action close to zero. This is exactly
what QM does: it gives the dynamics for small values of the action.
This is the reason why multiplying by $\frac{\theta\bar{\theta}}{\hbar}$
we get QM. Somehow this procedure projects out of \cite{KvN} the contribution
for small values of the action.

\section{Large values of the action}

In this section we will try to derive which is the theory for large
values of the action. Why do we do this? The reason is that those systems for which we have to invoke dark matter and dark energy are systems with very large values of the action. Think of  stars rotating very fast around the center of their galaxy, or of clusters of galaxies possessing huge masses and as a consequence a huge action or think of the universe when it started accelerating in its expansion 5 billions years ago.

We showed in the  previous  section  that $\theta\bar{\theta}$
has the dimension of an action. Multiplying the action by $\theta\bar{\theta}$
is the same as multiplying by $\delta\left(\theta\right)\delta\left(\bar{\theta}\right)$.
The product of these Dirac deltas is equivalent to $\delta\left(\theta\bar{\theta}\right)$
infact $\delta\left(\theta\bar{\theta}\right)=\delta\left(\theta\right)\frac{\partial}{\partial\theta}\delta\left(\theta\bar{\theta}\right)=\delta\left(\theta\right)\bar{\theta}=\theta\bar{\theta}$.
Now if we search for large values of the action, we should multiply
the action in (\ref{eq:def-generating-functional}) by $\delta\left(\frac{1}{\theta\bar{\theta}}\right)$.
We better pay attention in doing the inverse of a Grassmannian number
like $\theta\bar{\theta}$. As explained in ref.~\cite{dewitt-book} first we
should add a complex number $\epsilon$ to $\theta\bar{\theta}\rightarrow\epsilon+\theta\bar{\theta}$.
The inverse of $\epsilon+\theta\bar{\theta}$ is $\frac{1}{\epsilon}\left(1-\frac{\theta\bar{\theta}}{\epsilon}\right)$.
So in the action (\ref{eq:def-generating-functional}) we have to
insert
\begin{eqnarray}
 & \delta\left[\frac{1}{\epsilon}\left(1-\frac{\theta\bar{\theta}}{\epsilon}\right)\right]\,.\label{eq:3}
\end{eqnarray}
The Dirac delta (\ref{eq:3}) has its zero in $\theta\bar{\theta}=\epsilon$
and with $\epsilon\rightarrow0$, this implies $\theta\bar{\theta}=0$.
So for \emph{ large values of the action we have the same distribution as 
for  small values like in QM}. This is very surprising but at least it tells
us that the distribution is not the one of classical mechanics. In
fact where dark matter and dark energy are  invoked is because CM fails.

One last point: In the quantization procedure we had to divide the
$\theta\bar{\theta}$ by $\hbar$ in order to get QM. We wondered
by which factor we should devide in the case of large actions. It
has to have the dimension of the action but be very large. One such
factor could be $B=Mc^{2}T$ where $M$ is the mass of the universe
and $T$ the age of the universe. 
(This factor was suggested to me by Stefano Baroni.) 

We conclude this work with a question that we have not been able
to answer yet and which is the title of the paper: "are there problems
for large values of the action like there were for small ones". This is a natural question to ask as the two distributions are so similar 

\begin{acknowledgments}
\par This work has been supported
by the INFN of Trieste. I wish to thank C.Pagani for help with the latex.
This paper will not be sent to any journal but it will be waiting for comments, criticism and suggestions by the readers.
\end{acknowledgments}

\end{document}